\documentclass[12pt]{JHEP3}

\newcommand{\RR}{{ R}}
\newcommand{\pp}{$pp$-wave}
\renewcommand{\th}{\theta}
\newcommand{\beq}{\begin{equation}}
\newcommand{\eeq}{\end{equation}}
\newcommand{\bea}{\begin{eqnarray}}
\newcommand{\eea}{\end{eqnarray}}
\newcommand{\del}{\partial}
\renewcommand{\o}{\over}

\newcommand{\lap}{\Delta}

\renewcommand{\th}{\theta}

\newcommand{\om}{\omega}

\newcommand{\la}{\lambda}
\newcommand{\ri}{\right}
\newcommand{\lf}{\left}

\title{PP-Waves and Holography}

\author{Robert G. Leigh\\
       Department of Physics\\
       University of Illinois at Urbana-Champaign\\
       Urbana, IL 61801, USA\\
       E-mail: \email{rgleigh@uiuc.edu}}
\author{Kazumi Okuyama\\
       Enrico Fermi Institute\\
   University of Chicago\\
       Chicago, IL 60637, USA\\
       E-mail: \email{kazumi@theory.uchicago.edu}}
       \author{Moshe Rozali\\
       Department of Physics\\
   University of British Columbia\\
       Vancouver, BC, Canada\\
       E-mail: \email{rozali@physics.ubc.ca}}

\abstract{We consider aspects of holography in the $pp$-wave limit of
$AdS_5\times S^5$. This geometry contains two $\RR^4$'s, one obtained
from $AdS_5$ directions, and the other from the $S^5$. We argue that the
holographic direction in the $pp$-wave background can be taken to be
$r$, the radial direction in the first $\RR^4$. Normalizable modes
correspond to states, and non-normalizable modes correspond to
deformations of the boundary theory. In the strict $pp$-wave limit,
there are additional non-normalizable modes in the second $\RR^4$, which
have no apparent super-Yang-Mills interpretation. We outline the procedure for calculating correlation functions holographically.}


\preprint{ILL-(TH)-02-04\\ EFI-02-69}

\begin{document}

\section{Introduction}

Berenstein, Maldacena and Nastase\cite{bmn} have recently considered
\pp\ backgrounds from the point of view of the AdS/CFT correspondence.
Namely, the \pp\ is realized as a certain limit of $AdS_5\times S^5$,
where one considers a large boost along one of the $S^5$ directions. The
light cone string theory in this background contains eight massive
bosons and superpartners; it is exactly solvable, with a spectrum
\beq\label{eq:stringspec} H=\sum_{n=-\infty}^{\infty} N_n \sqrt{\mu^2+{4
n_{osc}^2\over(\alpha'p_-)^2}} \eeq BMN identified the oscillator modes
of this string within the dual $N=4$ SYM.

In addition to providing another example of a holographic dual,
the exciting new feature of \cite{bmn} is a more or less direct
connection between the boundary theory and the string theory in
this background, including non-BPS massive string modes. Examining
this duality can then provide a clue to the nature of holography
in a more generic setting. Previous work on \pp s include Refs.
\cite{prev}, and subsequent work includes \cite{orbifold, gravity,
bmn2,  string}.

Here we consider in more detail aspects of holography  for the
\pp. We would like to establish more clearly the holographic map.
To do so, we will concentrate in this paper, on the supergravity
modes (we are making an assumption that there is a decoupling
here, $\alpha'\mu p_->>1$). The map between normalizable
supergravity modes and states, and non-normalizable modes and
sources is well known in $AdS/CFT$. We explore this map in the
present context. In a sense, we may expect that this is determined
by the known $AdS_5\times S^5$ results, since the \pp\ geometry is
obtained in a scaling limit. On the other hand, questions such as
"where does the SYM theory live" are somewhat confusing. The \pp\
background has an $SO(4) \times SO(4)\subset SO(8)$ isometry; one
might well wonder what this means from the SYM point of view.

In this paper we take the point of view that describing the holographic
dual in terms of the original SYM theory amounts to making certain
choices that are arbitrary from the bulk viewpoint. This seems to choose
one of the two copies of $\RR^4$ in the geometry as the base space of
the SYM theory, and retains the memory of the  origin of the other copy
from a compact space (the original $S^5$). Perhaps this is the most
surprising aspect of our analysis- some aspects of the \pp\ background
(namely non-normalizable modes in directions originating from the
sphere) do not seem to be described by the original SYM theory.

Retaining features coming from the seemingly decoupled asymptotically
$AdS$ region, we formulate the holographic relation between this
background and $SYM$ theory. Up to some important differences we outline
below, the correspondence is the familiar one, with the radial
coordinate playing the role of the holographic one.

We discuss supergravity modes, normalizable and non-normalizable, in
the next section. We find that the spectrum of $p_-$ span the  positive real axis, and there exist the two types of modes for all such values. We conclude by  formulating the holographic calculation of correlation functions.  We hope to return to this calculation in the near future.

There are several interesting questions not answered by the present
work. The string theory in this background is simple, and one would hope
to be able to calculate more general amplitudes directly on the
worldsheet. However, presently the string theory involves a
Green-Schwarz formulation of the worldsheet theory, which is more
difficult to work with. It is also of interest to compare the gravity
(or string) calculations to the corresponding SYM calculations. This
would be a check of the holographic correspondence we suggest, and can
perhaps serve to formulate a holographic dual that is more directly
connected with the \pp background.

While this manuscript was in final stages of preparation, we
became aware of the preprints \cite{overlap}. These papers discuss
similar topics from a somewhat different viewpoint. In particular,
we emphasize here that as long as one uses the original SYM as the
boundary theory, the holographic coordinate which classifies
gravity modes is the radial one. Indeed, we find that this
classification simply descends from the corresponding
classification in the original $AdS$ space.

\section{Bulk Modes: Deformations and States}

The metric of the \pp\ is obtained through a suitable scaling
limit on  $AdS_5\times S^5$ in global coordinates \cite{bmn}. In
this coordinate system  the initial  background is dual to $N=4$
SYM theory on $\RR\times S^3$. The metric obtained in \cite{bmn}
is:
\begin{equation}\label{eq:metreight}
ds^2_{pp}=-4dx^+dx^--y_i^2\mu^2(dx^+)^2+dy_i^2
\end{equation}
where $i=1,\ldots,8$,
while the $RR$ background is
\begin{equation}
F_{+1234}=F_{+5678}=\mu
\end{equation}
This background is clearly $SO(4)\times SO(4)$ invariant; as such,
let us rewrite the metric in a suggestive form, in spherical
coordinates in each of the two $\RR^4$ factors.
 \begin{equation}\label{eq:metrfourfour}
ds^2_{pp}=-4dx^+dx^-+dr^2+r^2(d\Omega_3^2-\mu^2(dx^+)^2)
+d\tilde r^2+\tilde r^2(d\tilde \Omega_3^2-\mu^2(dx^+)^2)
\end{equation}
We note that at large $r$, we have an $S^3\times \RR$ coordinatized by
$\Omega_3, x^+$. Similarly, at large $\tilde r$, we also have an
$S^3\times \RR$ coordinatized by $\tilde\Omega_3, x^+$. In what follows,
we will show that the holographic coordinate is indeed $r$.
Equivalently, we could say that the holographic coordinate is $\tilde
r$; thus there are apparently two distinct "boundaries" in the \pp\
geometry. For the duality to SYM, we focus on one or the other.

Let us consider a scalar mode which is massless in ten
dimensions\footnote{Such a mode is dual to an operator which is a
descendant in the SYM theory, of dimension $\Delta = J +4$.}. In the
coordinates given in (\ref{eq:metreight}), the Laplacian is clearly
\beq
\Delta=-\partial_+\partial_-+{1\over4}\mu^2 y^2\partial_-^2+\Delta_8
\eeq
Solutions are of the form
\beq\label{eq:soleight}
\phi=e^{ip_+x^+}e^{ip_-x^-}\prod_{j=1}^8 e^{-\alpha_j y_j^2/4}
H_{n_j}\left(\sqrt{{\alpha_j\over 2}}y_j\right)
\end{equation}
with $p_+=\sum_{j}{\alpha_j\over p_-}(n_j+1/2)$, and $\alpha_j=\pm\mu
p_-$. For $\alpha>0$, the $H_n$'s are Hermite polynomials, and these
solutions are clearly normalizable.

As is familiar from $AdS/CFT$, we should not be so quick to discard the
non-normalizable solutions. Instead, we are faced with the task of
finding a criterion for distinguishing allowed modes from forbidden
ones\footnote{As this contradicts some statements in the literature, we demonstrate the existence of the non-normalizable modes in the appendix.}.

An analysis similar to the one leading to the  Breitenlohner-Freedman
bound suggests that we should take modes with $p_-$ positive only.
{}From the gauge theory side we have a variant of the familiar
light-cone quantization; it is a familiar aspect of lightcone treatment
that the spectrum of $p_-$ is semi-infinite. We claim this should be the
case both for (normalizable) states, and for (non-normalizable)
operators, or sources.

The bulk treatment clarifies the need for positivity of $p_-$.  In the
familiar $AdS$ story, there are two modes for each value of all quantum
numbers, one normalizable, and one non-normalizable\footnote{There is a
small window where both modes can be normalizable. This subtlety is
well-understood, and will play no role in our discussion.}. Those modes
have different values of the Hamiltonian (scaling dimension). They are
associated with each other since they carry the same quantum numbers;
thus turning on the non-normalizable mode will inevitably excite the
normalizable mode.

Now, for the normalizable modes the situation is clear- one should not
allow a normalizable state with a negative value of the Hamiltonian;
this would correspond to a runaway behavior of the vacuum. This is the
origin of the Breitenlohner-Freedman  bound in $AdS$ space \cite{bf}.
Excluding the normalizable state with a negative scaling dimension also
eliminates the corresponding source with the same quantum numbers.

In the present case, the Hamiltonian  is $p_+$, and one should not allow
normalizable modes with negative eigenvalues. This excludes all
normalizable modes with negative $p_-$.  The non-normalizable mode with
the same quantum numbers is also required to be absent. The quantum
numbers in question correspond to oscillator numbers in all eight
transverse directions, and to the light-cone momentum $p_-$.

To more fully appreciate the significance of (\ref{eq:soleight}), let us
consider the coordinatization given in (\ref{eq:metrfourfour}). The
scalar Laplacian is
\begin{equation}
\Delta=-\partial_+\partial_-+\left[{1\over 4}\mu^2 r^2\partial_-^2+
\Delta_{r,\Omega_3}\right]+\left[{1\over 4}\mu^2 \tilde r^2\partial_-^2+
\Delta_{\tilde r,\tilde\Omega_3}\right]
\end{equation}
Solutions can be written as
\begin{equation}
\phi=e^{ip_+x^+}e^{ip_-x^-}f(r,\Omega_3)\tilde f(\tilde r,\tilde\Omega_3)
\end{equation}
Acting on such modes, the Laplacian reduces to
\begin{equation}
\Delta=p_+p_-+\left[-{1\over 4}\mu^2p_-^2 r^2+\Delta_{r,\Omega_3}\right]+
\left[-{1\over 4}\mu^2p_-^2 \tilde r^2+\Delta_{\tilde r,\tilde\Omega_3}\right]
\end{equation}
Any given solution will satisfy
\begin{eqnarray}
\Delta_{\tilde r,\tilde\Omega_3}\tilde f-{1\over 4}\mu^2p_-^2 \tilde r^2\tilde f
=-\tilde E\tilde f\\
\Delta_{r,\Omega_3} f-{1\over 4}\mu^2p_-^2  r^2 f=- E f
\end{eqnarray}
with
\begin{equation}
p_+={E+\tilde E\over p_-}
\end{equation}
This form is consistent with the $n_{osc}=0$ part of eq. (\ref{eq:stringspec}).

Solutions are readily obtained as
\begin{equation}
f=e^{-\alpha r^2/4} r^{\beta}Y_{\ell, m_1, m_2}(\Omega_3)
\end{equation}
with a similar expression for $\tilde f$. If we want the solutions
to be singularity-free in the interior, $r\sim 0$, then
$\beta=\ell$. The differential equation then reduces to
\begin{equation}
E=\alpha(\ell+2)
\end{equation}
For each $\ell$, there is a certain degeneracy, which is accounted for
in the  analysis in Cartesian coordinates (see the appendix for details).

So the analysis of the gravity modes suggests that it is sensible to
formulate question in the SYM theory regarding modes with positive
$p_-$, and to answer those questions holographically. The gravity modes
are divided in the familiar way: normalizable modes are states in the
Hilbert space of the theory, and non-normalizable modes are sources
(deformations) of the theory.  We proceed next to outline the
procedure to calculate the correlation functions of arbitrary modes of
positive $p_-$.

The picture is less clear regarding  the other $\RR^4$ contained in the
\pp\ geometry.  {\it A priori}, in considering propagation on the \pp\
background, one should consider both normalizable and non-normalizable
modes in that $\RR^4$ as well.  It is not clear to us, though, that the
SYM formulation knows about both types of modes.

To clarify this statement, let us derive the modes in this background as
limits of modes in the original $AdS_5 \times S^5$ background. Let us
consider the scalar field $\phi$ on $AdS_5$ with metric
\beq ds^2_{AdS_5}=R^2(-\cosh^2\rho dt^2
+d\rho^2+\sinh^2\rho d\Omega_3^2).
 \eeq
The Laplacian is
\beq
\lap={1\o R^2}\lf(-{1\o\cosh^2\rho}\del_t^2-{\ell(\ell+2)\o\sinh^2\rho}
+{1\o\cosh\rho\sinh^3\rho}\del_{\rho}(\cosh\rho\sinh^3\rho\del_{\rho})\ri).
\eeq

The simplest solution of $(\lap-m^2)\phi=0$ is given by (see review
\cite{bigrev}, eq. (2.34))\footnote{Note that
$\sin\th=\tanh\rho,\cos\th=1/\cosh\rho$ \cite{bigrev}. 
This solution corresponds to
the case $n=0$ in eq.(2.41) of \cite{bigrev}, 
{\it i.e.}, ${}_2F_1(a,b,c;\tanh\rho)=1$.
In the general case $n\not=0$, we have a factor
${}_2F_1(a,b,c;\tanh\rho)$. However, this factor can be neglected in
the limit $R\rightarrow\infty$ since
${}_2F_1(a,b,c;\tanh(r/R))\rightarrow 1$. }
\beq
\phi=e^{i\omega
t}(\cosh\rho)^{-\la}(\tanh\rho)^\ell Y_{\ell,m_1,m_2}(\Omega_3),\quad 
(\om R)^2=(\la+\ell)^2.
\eeq
In the $AdS_5$ case, $\la$ is given by
\beq
\la=\lap_{\pm}=2\pm\sqrt{4+m^2R^2}.
\eeq
In the limit
$\lap\equiv\lap_+\sim R^2\sim\sqrt{N}>>1$, the normalizable mode
$\lap_+$ corresponds to $\la=\lap$ and the non-normalizable mode
$\lap_-=4-\lap_+\sim-\lap$ corresponds to $\la=-\lap$.

In the scaling limit
\beq
\rho={r\o R},\quad
\la=\pm\lap=\pm{1\o2}(p_++p_-R^2),\quad R\rightarrow\infty,
\eeq
with $r, p_+=\lap-J, p_-=(\lap+J)/R^2>0$ fixed, this $AdS$ solution
reduces to
\bea
\phi &\rightarrow& e^{i\om
t}\lf(1+{r^2\o2R^2}\ri)^{\mp{1\o2}p_-R^2}
\lf({r\o R}\ri)^{\ell}Y_{\ell,m_1,m_2}(\Omega_3)\nonumber \\
&\sim& e^{i\om t}e^{\mp{1\o4}p_-r^2}r^\ell Y_{\ell,m_1,m_2}(\Omega_3).
\eea

One can introduce the parameter $\mu$ by the redefinition
$x^{\pm}\rightarrow \mu^{\pm} x^{\pm}$ as usual. Then $p_{\pm}$
goes to $\mu^{\mp}p_{\pm}$. Therefore, these solutions agree with
those we obtained in the \pp\ background. Moreover, the
normalizability in the \pp\  limit,  {\it i.e.,} the sign of the Gaussian
factor descends from the choice of $\lap_+$ or $\lap_-$.

Therefore we see that by scaling of the modes on $AdS_5 \times
S^5$ we obtain both types of modes in the $AdS$ directions.
However, we started with normalizable modes only in the sphere
directions. The non-normalizable modes in directions originating
form the sphere are then not expected to be included in the
original SYM description.

 The situation is analogous to the
original derivation of $AdS/CFT$ from the near horizon geometry of
the 3-brane. One obtains $AdS_5 \times S^5$ in Poincar\'e
coordinates. In the decoupling limit a lot of the original
structure of the asymptotically flat space is eliminated, of
course. In our case the analogous  statement is that most of the
structure of the SYM theory will be irrelevant in the limit taken.
This is apparent, for example, in the statement that most of the
virtual processes appearing in SYM diagrams will have vanishing
contributions to the amplitudes we are interested in.

However, in the $AdS$ case, there were also additional structures
in the theory which are only visible in global coordinates. This is
visible only when one truly eliminates even the possibility of
appending an asymptotically flat region to the geometry.

Therefore, it seems to us that insisting on describing the system
by the {\it original} holographic dual retains some features of
the original asymptotic region. It may well be the case that this
eliminates the possibility of discussing non-normalizable modes in
the other directions. Such discussion naively requires a higher
dimesnional holographic dual. Since  a putative dual is absent, we
proceed by assuming that the modes are normalizable in the
directions originating from the sphere.

\section{Conclusions-Correlation Functions}

The next step in discussing the radial holography is a computation
of the correlation functions. We sketch here the procedure for
such a calculation.

The holographic calculation of the partition function with
arbitrary sources is performed by fixing a surface near the
boundary, and solving the Dirichlet problem,  finding the fields
with fixed sources on the surface. The partition function is then,
in the supergravity approximation, the value of the action on
shell.

A primary role in this calculation is played by the bulk to
boundary propagator. The solution to the Dirichlet problem is the
propagator convoluted with the given boundary source.  In $AdS_5
\times S^5 $ the bulk to boundary operator $K_{m^2}(x,r;y)$ is
defined for each mode of mass $m$, where $x,y$ are boundary
points, and $r$ is the holographic coordinate (so that $x,r$
specifies a bulk position). It is required to satisfy the Laplace
equation: \beq (\partial^2 - m^2) K_{m^2} =0\eeq with the boundary
conditions \beq
 K(x,r;y) \rightarrow r^\Delta \delta(x-y) ~~~~~~{\rm as}\ r\ {\rm approaches\
 the\ boundary}
 \eeq
 where $\Delta$ is determined by the behavior of the
 non-normalizable mode at the boundary- it depends only on $m^2$.

 One way of solving the equation is as follows. Consider a
 complete set of functions on the boundary, $\psi_n(x)$, and modes
 of the bulk Laplacian that satisfy
 \beq
 \Phi_n (r,x)\rightarrow r^\Delta \psi_n(x) ~~~~~~~{\rm as}\ r\ {\rm approaches\
 the\ boundary}
 \eeq

 We note that the constant $\Delta$ is required to be identical for
 all modes, independent of $n$. It is uniquely specified by the mass $m$.
 In this case it is easy to check
 that
 \beq
K(x,r;y) = \sum_n \Phi_n (r,x)\psi_n(y) \eeq

This is the bulk to boundary propagator. Since the set of boundary
sources $\psi_n$ is complete, one is able to turn on an arbitrary
boundary source for any operator of a definite scaling dimension
(which is related to $m$). The natural sources to consider then
have definite values  of the $SO(6)$ Casimir ($m^2$).

Similar procedure can be obtained in our case. A complete set of
functions on the boundary, which is of the form $R \times S^3$,
can be chosen to carry definite $p_+$ and $SO(4)$ quantum numbers.
We have then $\psi_n(x_+, \Omega_3)= e^{i p_+ X_+} Y_I
(\Omega_3)$. Here the angles on the sphere are denoted by
$\Omega_3$, and $Y_I (\Omega_3)$ are scalar spherical harmonics.

The source function on the boundary determines the behavior of the
corresponding non-normalizable modes. The internal quantum numbers
in our case are $p_-$, and the angular and radial quantum numbers
in the additional copy of $\RR^4$. The behavior of the
non-normalizable mode only depends on $p_-$, so one has to work in
a definite $p_-$ basis. The quantum number $p_-$ plays a role
similar to the Casimir $m^2$ in the $AdS_5 \times S^5$ case.

A slight complication here is the absence of coordinate system in
which the consequences of conformal invariance are transparent. In
$AdS$ one usually works in Poincare coordinates, where a surface
near the boundary corresponds to an ultraviolet cutoff which
manifests cleanly the consequences of conformal invariance. Here
we are restricted to work in global coordinates, where the \pp
limit is done, so we are led to a more complicated procedure. We
hope to report on progress in this direction in the near future.

\bigskip
\noindent {\bf Acknowledgments}: We thank David Berenstein and
Gordon Semenoff for useful conversations. RGL and KO thank the
University of British Columbia for hospitality. Work supported in
part by US DOE grant DE-FG02-91ER40677.

\section*
{Appendix: Four Dimensional Harmonic Oscillator}

We work out a simple quantum mechanics problem, considering the
eigenmodes of a four dimensional harmonic oscillator. The
questions we are interested in are naturally different from the
conventional ones, for example the existence of non-normalizable
modes and their behavior at radial infinity.

In Cartezian coordinates the problem is readily seperable to four
identical harmonic oscillators, satisfying

\beq f^{''} -\mu^2 {p_-}^2 x^2 \, f(x) = E\, f(x) \eeq

We can then write $f=e^{-\frac{1}{2} \alpha x^2} g(x)$. One gets:

\beq g^{''} - 2\alpha x g^{'} +(E -\alpha)g =0 \eeq

where we choose $\alpha = \pm \mu p_-$. We note that we reserve
the choice of $\alpha$ being positive or negative, corresponding
to normalizable or non-normalizable modes.

The equation for $g(x)$ is almost identical to the familiar
Hermite equation. We need to rescale the coordinate $x= ay$ to get
to the form:

\beq g^{''} - 2 y  g^{'} + 2n g =0 \eeq

For this to be correct one has to choose $a^2 =\alpha$, so
$\alpha$ has to be positive. In that case one obtains the equation
for the Hermite polynomials, provided:

\beq E_n = \alpha (n+\frac{1}{2}) \eeq

This yields  normalizable solution for any sign of $p_-$. One
simply has to choose $\alpha = \mu |p_-|$.

 A different  solution is obtained by  choosing $\alpha = -\mu |p_-|$,
  which
yields an exponentially growing mode. One can still do the above
change of variables for $a^2 =- \alpha$. One gets a slightly
different equation for $g$

\beq
 -g^{''} - 2 y  g^{'} + (\frac{E}{\alpha} -1)  g =0
 \eeq

The solutions of this equation are polynomials of the form $P_n =
i^n H_n(iy)$. These are real polynomials in y (the imaginary
normalization factors for $n$ odd do not matter  for a linear
equation). They modes we obtain are simply polynomials multiplied
by an exponentially growing Gaussian. In particular they do not
blow up at the origin.

The spectrum of $E$ is the same for those non-normalizable modes.
Note however that $\alpha$ is negative here, so the spectrum is
negative definite.

\end{document}